\documentclass[conference]{IEEEtran}
\IEEEoverridecommandlockouts

\usepackage{cite}
\usepackage{amsmath,amssymb,amsfonts}
\usepackage{graphicx}
\usepackage{textcomp}
\usepackage{booktabs}
\usepackage{multirow}
\usepackage{makecell}
\usepackage{tabularx}
\usepackage{url}
\usepackage{balance}
\usepackage{microtype}
\usepackage[hidelinks]{hyperref}

\begin{document}

\title{Attention-Enhanced Dual-Branch ConvNeXt--BiLSTM Network for Subject-Independent EEG Seizure Detection}

\author{
\IEEEauthorblockN{1\textsuperscript{st} Maimuna Chowdhury}
\IEEEauthorblockA{\textit{Dept. of Computer Science and Engineering} \\
\textit{Khulna University of Engineering \& Technology}\\
Khulna, Bangladesh \\
\href{mailto:maimuna.cse.kuet@gmail.com}{maimuna.cse.kuet@gmail.com}}
\and
\IEEEauthorblockN{2\textsuperscript{nd} Dr. Sk. Imran Hossain}
\IEEEauthorblockA{\textit{Dept. of Computer Science and Engineering} \\
\textit{Khulna University of Engineering \& Technology}\\
Khulna, Bangladesh \\
\href{mailto:supervisor@kuet.ac.bd}{imran@cse.kuet.ac.bd}}
}

\maketitle

\begin{abstract}
Automated seizure detection from scalp electroencephalography (EEG) is difficult because seizure morphology varies among patients and seizure samples are substantially outnumbered by non-seizure samples. This paper presents an attention-enhanced dual-branch network that jointly learns time--frequency and temporal representations from the same EEG segment. A continuous wavelet transform converts each segment into a scalogram processed by an ImageNet-pretrained ConvNeXt-Tiny backbone and squeeze-and-excitation attention. In parallel, a bidirectional long short-term memory network followed by multi-head self-attention models the raw signal. The two feature vectors are concatenated and classified by a weighted multilayer perceptron. Experiments use 14 subjects from the CHB-MIT scalp EEG database with subject-wise partitioning performed before overlapping segmentation. The model obtains $97.88\%$ accuracy and $97.51\%$ F1-score over ten across-subject splits, and $97.51\%$ accuracy, $96.59\%$ F1-score, and $98.02\%$ area under the ROC curve under 14-fold leave-one-subject-out validation. Removing temporal attention causes the largest ablation loss. The model requires 28.26 million parameters and 4.56 GFLOPs, with a measured network-only inference latency of 4.64 ms per segment.
\end{abstract}


\section{Introduction}
Epilepsy is characterized by recurrent seizures associated with abnormal neuronal activity. Scalp electroencephalography (EEG) is central to seizure assessment because it captures electrical brain activity non-invasively and at high temporal resolution. Manual inspection of long-duration recordings is, however, labor-intensive and susceptible to inter-observer variability. Automated systems can flag candidate seizure intervals for expert review, but building one that works reliably across patients is hard: seizure duration, frequency content, spatial spread, and morphology all vary from person to person \cite{bhattacharyya2017multivariate,li2021seizure}.

Early detection pipelines paired hand-engineered time-, frequency-, or time--frequency descriptors with classifiers such as support vector machines and random forests \cite{subasi2010eeg,mursalin2017automated}. Deep networks removed much of this manual feature design by learning representations straight from the EEG: one-dimensional convolutional and recurrent models pick up local and temporal patterns, while two-dimensional convolutional models work on spectrograms or wavelet scalograms instead \cite{zhao2020seizurenet,zhao2024residual}. Combining both input types is appealing for this reason -- the raw waveform keeps temporal detail intact, and a time--frequency view surfaces transient spectral changes that are harder to see in the original sequence \cite{tian2019deep,pan2022epileptic}.

Three practical gaps remain in existing hybrid designs. Many of them lean on either the raw sequence or the image representation, without refining both branches with attention. Reported performance often comes from random sample-wise splits, which can leak correlated windows from the same patient or recording into both training and test sets. And a high segment-level accuracy alone says little about clinical readiness -- that requires subject-independent evaluation and a look at computational cost.

To address these gaps, we build a dual-branch network: one ConvNeXt branch learns spatial--spectral features from continuous-wavelet scalograms, and a bidirectional long short-term memory (BiLSTM) branch learns temporal features directly from the raw EEG. Squeeze-and-excitation (SE) attention recalibrates the ConvNeXt channels, and multi-head self-attention (MHSA) highlights the most informative temporal states. Concretely, this paper contributes the following:
\begin{itemize}
    \item a heterogeneous fusion model combining raw EEG and continuous-wavelet scalograms, with branch-specific attention mechanisms;
    \item subject-wise evaluation using both repeated across-subject splits and 14-fold leave-one-subject-out (LOSO) validation, with partitioning before overlapping segmentation; and
    \item controlled baseline, ablation, and computational analyses that isolate the value and cost of each architectural component.
\end{itemize}

\section{Related Work}
Traditional seizure detectors employ statistical, spectral, wavelet, entropy, or autoregressive features followed by a conventional classifier \cite{gotman1982automatic,park2011seizure,zhang2017ar}. Although such pipelines can be efficient, their performance depends strongly on feature design and may not generalize when seizure morphology changes across subjects.

Deep-learning approaches learn features automatically. SeizureNet used convolutional processing for robust EEG seizure detection \cite{zhao2020seizurenet}, while residual and recurrent combinations have been used to capture local and long-range dependencies \cite{zhao2024residual}. Song \emph{et al.} combined dynamic channel screening, short-time Fourier transform (STFT), and a ResNet--LSTM model, reporting $96.59\%$ across-subject accuracy \cite{song2024optimization}. Abdulwahhab \emph{et al.} used STFT images and raw EEG in a CNN--LSTM architecture and reported $97.12\%$ accuracy under a hold-out protocol \cite{abdulwahhab2024detection}. Shen \emph{et al.} compared STFT and continuous wavelet transform (CWT) images with GoogLeNet under LOSO evaluation \cite{shen2024real}. More recently, Zhao \emph{et al.} combined an S-transform, ConvNeXt, and SimAM attention, although the reported $98.83\%$ accuracy was obtained using a fixed 8:1:1 split rather than subject-independent cross-validation \cite{zhao2025advanced}.

CWT is suitable for non-stationary EEG because its scale-dependent windows provide variable time--frequency resolution. ConvNeXt modernizes convolutional design using depthwise convolution, large kernels, inverted bottlenecks, layer normalization, and GELU activation \cite{liu2022convnet}. The present study combines these properties with a separate raw-signal branch and evaluates whether SE and temporal self-attention provide complementary gains.

\section{Materials and Methods}
\subsection{Dataset and Subject Selection}
All experiments draw on the CHB-MIT scalp EEG database, a public dataset distributed via PhysioNet \cite{goldberger2000physiobank}. It holds long-term pediatric recordings captured at a 256 Hz sampling rate, with seizure intervals marked by clinical experts, and electrodes placed according to the international 10--20 system, mostly as bipolar derivations.

Fourteen subjects with a compatible common-channel configuration were selected: CHB01, CHB03, CHB06, CHB08, CHB10, CHB11, CHB14, CHB18, CHB19, CHB20, CHB21, CHB22, CHB23, and CHB24. These subjects contain 91 annotated seizure events totaling 5,429 s of seizure activity. Only bipolar channels shared by the selected records were retained. Files or subjects with incompatible channel configurations were excluded so that every model input followed the same channel definition.

\subsection{Preprocessing and Segmentation}
We suppress baseline drift and high-frequency noise with a fourth-order Butterworth band-pass filter (0.1--50 Hz cutoffs), applied independently to each retained bipolar channel as a one-dimensional signal. Each filtered channel is then cut into 5-s windows at $50\%$ overlap -- 1,280 samples per window with a 640-sample stride -- and labeled seizure or non-seizure from the database annotations. Because overlapping windows from the same subject are correlated, subjects are assigned to train or test before this segmentation step, not after.

Channel-wise $z$-score normalization reduces amplitude variation across recordings. Training-only augmentation applies Gaussian noise ($p=0.30$), temporal shifting ($p=0.25$), amplitude scaling ($p=0.20$), and frequency-domain perturbation ($p=0.15$). Class imbalance is handled jointly through a weighted random sampler and positive-class weighting in the binary cross-entropy loss.

\subsection{Continuous-Wavelet Scalogram}
For a signal $x(t)$, the CWT coefficient at scale $a$ and translation $b$ is
\begin{equation}
W_x(a,b)=\frac{1}{\sqrt{|a|}}\int_{-\infty}^{\infty}x(t)\psi^{*}\left(\frac{t-b}{a}\right)dt,
\end{equation}
where $\psi$ is the mother wavelet. A Morlet wavelet is used because it provides localized oscillatory analysis. The coefficient magnitudes are log-transformed, clipped at the 5th and 95th percentiles, and normalized. Each scalogram is resized to $224\times224$, mapped to three channels with the Viridis colormap, and normalized using ImageNet statistics.

\subsection{Attention-Enhanced Dual-Branch Network}
\begin{figure*}[t]
    \centering
    \includegraphics[width=0.94\textwidth]{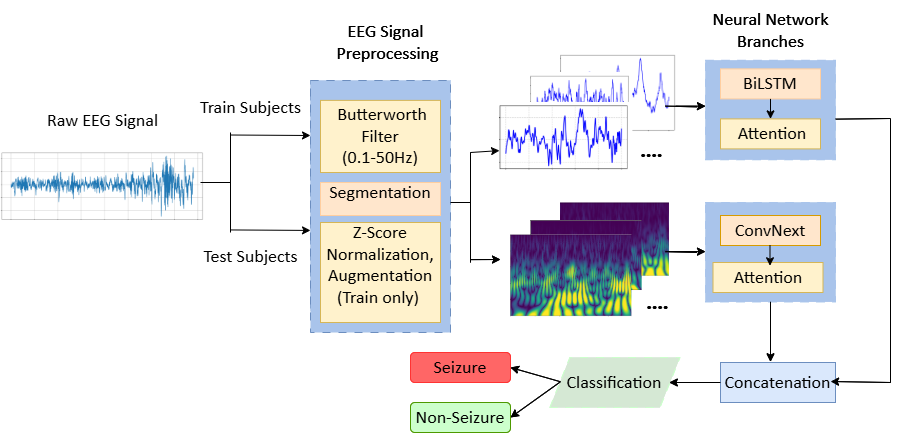}
    \caption{Proposed pipeline. Each filtered EEG window is represented simultaneously as a raw sequence and a CWT scalogram. The two attention-refined branches are concatenated before binary classification.}
    \label{fig:architecture}
\end{figure*}

Fig.~\ref{fig:architecture} summarizes the complete pipeline. The upper branch receives the raw 1,280-sample sequence. A one-layer BiLSTM with 64 hidden units in each direction produces a 128-dimensional sequence representation. Four-head self-attention then relates every temporal state to all other states. For input states $H$, one attention head is computed as
\begin{equation}
\operatorname{Attn}(Q,K,V)=\operatorname{softmax}\left(\frac{QK^{\mathsf T}}{\sqrt{d_h}}\right)V,
\end{equation}
where $Q=HW_Q$, $K=HW_K$, $V=HW_V$, and $d_h$ is the head dimension. The four head outputs are concatenated and projected to a 128-dimensional temporal feature vector.

The lower branch processes the CWT scalogram using an ImageNet-pretrained ConvNeXt-Tiny backbone. The backbone produces a 768-dimensional spatial--spectral descriptor. An SE block first aggregates spatial information by global average pooling and then learns channel weights through a two-layer bottleneck with reduction ratio 16. Multiplication by these weights emphasizes informative ConvNeXt channels and suppresses less useful responses.

The 768-dimensional scalogram vector and 128-dimensional temporal vector are concatenated into an 896-dimensional fused representation. A multilayer perceptron with dimensions $896\rightarrow256\rightarrow128\rightarrow1$, ReLU hidden activations, dropout, and a sigmoid decision layer predicts seizure probability.

\subsection{Training and Evaluation Protocols}
The network is trained in PyTorch for at most 30 epochs using AdamW, weight decay 0.05, batch size 64, gradient clipping at 1.0, mixed-precision computation, and a ReduceLROnPlateau scheduler. Layer-wise learning rates range from $5\times10^{-8}$ to $8\times10^{-5}$. The random seed is fixed at 42.

Two subject-independent protocols are used. In the repeated across-subject experiment, 11 subjects are used for training and three unseen subjects for testing; ten different subject combinations are evaluated. In LOSO validation, one subject is held out for testing and the remaining 13 subjects are used for training, yielding 14 folds. Accuracy, precision, sensitivity, specificity, F1-score, and area under the receiver operating characteristic curve (AUC) are calculated at the segment level. Results are reported as mean and standard deviation across splits or folds.

\section{Results and Discussion}
\subsection{Overall Performance}
Table~\ref{tab:overall} summarizes both protocols. The repeated across-subject splits are tight, with accuracy varying by only 0.52 percentage points. LOSO is a harder test -- each patient is evaluated in isolation, so the spread is wider -- yet mean accuracy still holds at $97.51\%$ and mean F1-score at $96.59\%$. AUC stays high in both settings ($98.52\%$ and $98.02\%$), showing the model separates the two classes well across thresholds. That said, LOSO's lower precision and F1-score compared to the repeated split confirm that patient-specific variability is still a real source of error, not just noise.

\begin{table}[t]
\caption{Subject-Independent Detection Performance (Mean $\pm$ Standard Deviation, \%)}
\label{tab:overall}
\centering
\resizebox{\columnwidth}{!}{%
\begin{tabular}{lcccccc}
\toprule
Protocol & Acc. & Prec. & Sens. & Spec. & F1 & AUC \\
\midrule
Across-subject & $97.88\pm0.52$ & $97.32\pm0.57$ & $97.70\pm0.43$ & $98.12\pm0.42$ & $97.51\pm0.49$ & $98.52\pm0.30$ \\
LOSO & $97.51\pm2.01$ & $96.21\pm2.11$ & $96.97\pm2.05$ & $97.76\pm1.93$ & $96.59\pm2.08$ & $98.02\pm1.84$ \\
\bottomrule
\end{tabular}}
\end{table}

The full subject-wise LOSO breakdown is in Table~\ref{tab:loso_subjects}. CHB06 and CHB14 are the hardest subjects to hold out, while several others clear $99\%$ accuracy -- this spread is exactly why the LOSO standard deviation is larger, and why we report per-subject numbers rather than only the aggregate.

\begin{table*}[t]
\caption{Complete Subject-Wise LOSO Results (\%)}
\label{tab:loso_subjects}
\centering
\scriptsize
\begin{tabular}{lcccccc}
\toprule
Test subject & Accuracy & Precision & Sensitivity & Specificity & F1-score & AUC \\
\midrule
CHB01 & 97.89 & 96.67 & 97.45 & 98.45 & 97.06 & 98.67 \\
CHB03 & 99.23 & 98.12 & 98.78 & 99.56 & 98.45 & 99.67 \\
CHB06 & 93.45 & 91.89 & 92.78 & 94.23 & 92.33 & 94.89 \\
CHB08 & 99.45 & 98.34 & 99.01 & 99.78 & 98.67 & 99.89 \\
CHB10 & 98.12 & 96.89 & 97.67 & 98.78 & 97.28 & 98.89 \\
CHB11 & 95.67 & 94.12 & 95.01 & 96.34 & 94.56 & 96.78 \\
CHB14 & 93.89 & 92.34 & 93.23 & 94.67 & 92.78 & 95.23 \\
CHB18 & 98.78 & 97.56 & 98.23 & 99.12 & 97.89 & 99.23 \\
CHB19 & 96.34 & 94.89 & 95.78 & 97.01 & 95.33 & 97.45 \\
CHB20 & 99.01 & 97.89 & 98.56 & 99.34 & 98.22 & 99.45 \\
CHB21 & 96.78 & 95.34 & 96.23 & 97.45 & 95.78 & 97.89 \\
CHB22 & 99.34 & 98.23 & 98.89 & 99.67 & 98.56 & 99.78 \\
CHB23 & 98.45 & 97.23 & 97.89 & 99.01 & 97.56 & 99.12 \\
CHB24 & 98.67 & 97.45 & 98.12 & 99.23 & 97.78 & 99.34 \\
\midrule
Mean $\pm$ SD & $97.51\pm2.01$ & $96.21\pm2.11$ & $96.97\pm2.05$ & $97.76\pm1.93$ & $96.59\pm2.08$ & $98.02\pm1.84$ \\
\bottomrule
\end{tabular}
\end{table*}

\begin{figure}[t]
    \centering
    \includegraphics[width=\columnwidth]{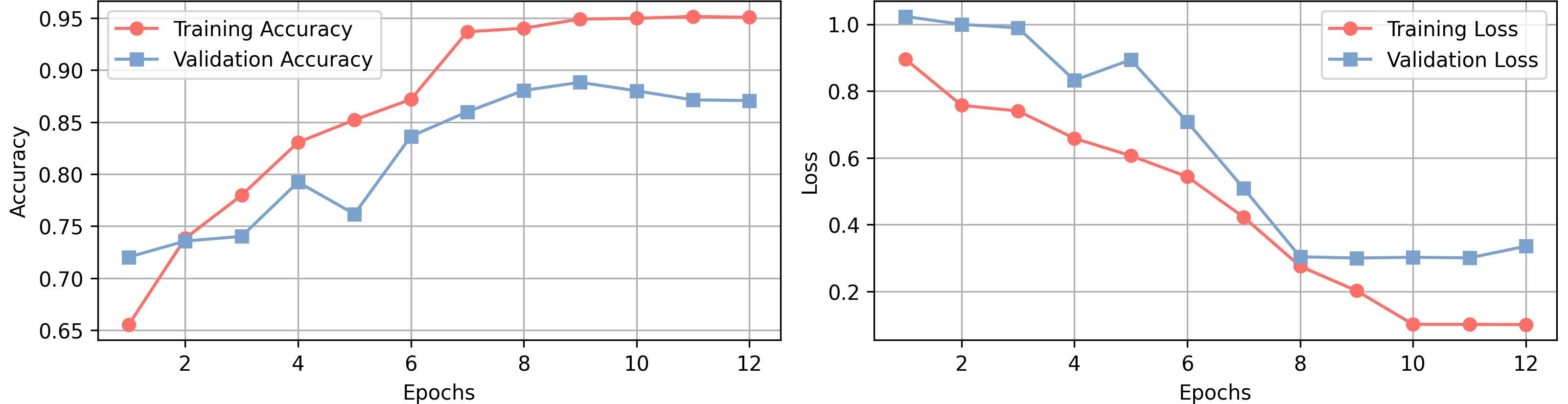}
    \caption{Training and validation accuracy (left) and loss (right) for a representative LOSO fold.}
    \label{fig:curves}
\end{figure}

Fig.~\ref{fig:curves} shows training and validation behavior for one representative LOSO fold, which converges stably. We include this only as diagnostic evidence of stable optimization -- the cross-subject test metrics above remain the actual measure of generalization.

\subsection{Baselines and Ablation Analysis}
Table~\ref{tab:baseline} compares the proposed method against models trained under the same preprocessing and repeated across-subject protocol. The dual-input network beats the ConvNeXt-only baseline by 4.22 points in accuracy and the BiLSTM baseline by 9.81 points, evidence that the fusion is doing real work: the scalogram branch captures spectral morphology the raw sequence misses, and the temporal branch keeps ordering information that can get diluted when the signal is converted to an image.

The ablation results in Table~\ref{tab:ablation} separate the attention mechanisms. The full network performs best. Retaining MHSA while removing SE gives $96.40\%$ accuracy; retaining SE while removing MHSA gives $94.99\%$. Thus, removal of temporal self-attention causes the larger decrease, suggesting that global temporal interactions are particularly important for distinguishing seizure and non-seizure windows. Removing both mechanisms reduces accuracy to $94.08\%$.

\begin{table}[t]
\caption{Baseline Models Under the Same Across-Subject Protocol}
\label{tab:baseline}
\centering
\begin{tabular}{lccc}
\toprule
Model & Input & Accuracy (\%) & F1 (\%) \\
\midrule
CNN & Scalogram & 84.21 & 84.09 \\
ResNet50 & Scalogram & 92.58 & 93.71 \\
BiLSTM & Raw EEG & 88.07 & 90.86 \\
ConvNeXt-only & Scalogram & 93.66 & 93.21 \\
\textbf{Proposed} & Both & \textbf{97.88} & \textbf{97.51} \\
\bottomrule
\end{tabular}
\end{table}

\begin{table}[t]
\caption{Attention Ablation Under the Across-Subject Protocol}
\label{tab:ablation}
\centering
\begin{tabular}{lcc}
\toprule
Variant & Accuracy (\%) & F1 (\%) \\
\midrule
Full model (SE + MHSA) & \textbf{97.88} & \textbf{97.51} \\
MHSA only (without SE) & 96.40 & 95.33 \\
SE only (without MHSA) & 94.99 & 94.76 \\
Without SE and MHSA & 94.08 & 93.97 \\
\bottomrule
\end{tabular}
\end{table}

\subsection{Comparison with Previously Reported Results}
Table~\ref{tab:comparison} places the results alongside representative CHB-MIT studies. Such comparisons must be interpreted cautiously because subject selection, channel configuration, window construction, balancing, and validation differ. The proposed method is competitive with methods evaluated under subject-independent settings. The higher result reported by the S-transform--ConvNeXt method used a fixed train/validation/test split and is therefore not directly equivalent to LOSO evaluation.

\begin{table}[t]
\caption{Comparison with Representative CHB-MIT Studies}
\label{tab:comparison}
\centering
\resizebox{\columnwidth}{!}{%
\begin{tabular}{llccc}
\toprule
Study & Technique & Protocol & Acc. (\%) & F1 (\%) \\
\midrule
Shen \emph{et al.} \cite{shen2024real} & CWT + GoogLeNet & LOSO & 93.20 & -- \\
Shen \emph{et al.} \cite{shen2024real} & STFT + GoogLeNet & LOSO & 97.74 & -- \\
Song \emph{et al.} \cite{song2024optimization} & STFT + ResCon-LSTM & Across & 96.59 & 96.37 \\
Abdulwahhab \emph{et al.} \cite{abdulwahhab2024detection} & STFT + CNN-LSTM & Hold-out & 97.12 & 97.27 \\
Zhao \emph{et al.} \cite{zhao2025advanced} & S-trans. + ConvNeXt & 8:1:1 & 98.83 & -- \\
\textbf{Proposed} & CWT + raw EEG & LOSO & 97.51 & 96.59 \\
\textbf{Proposed} & CWT + raw EEG & Across & 97.88 & 97.51 \\
\bottomrule
\end{tabular}}
\end{table}

\subsection{Computational Cost and Limitations}
The complete network contains 28.26 million parameters and requires approximately 4.56 GFLOPs per input. ConvNeXt dominates the cost with 27.82 million parameters and 4.47 GFLOPs. BiLSTM, SE, MHSA, and the classifier collectively add less than 0.45 million parameters. On the experimental NVIDIA GPU, the trained network processes a segment in 4.64 ms, corresponding to 215.6 segments/s and approximately 115 MB of measured inference memory.

The latency value covers only the forward pass. It excludes filtering, segmentation, CWT generation, image resizing, host-to-device transfer, and post-processing; therefore, it should not be interpreted as end-to-end clinical alarm latency. Additional limitations are that only 14 CHB-MIT subjects were included to obtain a common montage, performance was measured at the segment rather than seizure-event level, and no external dataset was used. Overlapping windows increase the number of samples but also increase correlation within a subject; subject-wise partitioning prevents direct leakage but does not remove this within-training redundancy. Future evaluation should report event sensitivity, false alarms per hour, detection delay, calibration, confidence intervals or paired significance tests, and external cross-dataset performance.

\section{Conclusion}
This paper presents an attention-enhanced dual-branch network for EEG seizure detection. A ConvNeXt--SE branch learns spatial--spectral structure from CWT scalograms, while a BiLSTM--MHSA branch models raw temporal dynamics. Their fusion achieves $97.88\%$ accuracy in repeated across-subject experiments and $97.51\%$ accuracy under 14-fold LOSO validation. Baseline and ablation studies show that heterogeneous fusion improves performance and that temporal self-attention contributes the larger attention-related gain. These results are encouraging, but the model is not ready for deployment -- external validation and event-level clinical metrics are still needed. We plan to explore lighter backbones, streaming CWT computation, and channel- and time-specific interpretability in future work.

\balance
\bibliographystyle{IEEEtran}
\bibliography{references}

\end{document}